\documentclass[amstex,thmsa,sw20aip]{article}
\usepackage{amssymb}
\usepackage{amsfonts}
\usepackage{amsmath}

\begin{document}

\begin{center}
{\large Charged Brownian Particles: }

{\large Kramers and Smoluchowski Equations }

{\large and the Hydrothermodynamical Picture. }

{\large \vspace{0.3cm}}

R. E. Lagos$^{a)}$ and Tania P. Sim\~{o}es$^{b)}$

$^{a)}$ Departamento de F\'{\i}sica, IGCE, UNESP (Universidade Estadual
Paulista)\\[0pt]
CP. $178$, $13500$-$970$ Rio Claro SP Brazil.\\[0pt]
$^{b)}$ Instituto de F\'{\i}sica \textit{Gleb Wataghin }UNICAMP\
(Universidade Estadual de Campinas) \\[0pt]
CP. $6165$, $13083-970$ Campinas, SP Brazil.\\[0pt]
\end{center}

\vspace{1cm}

We consider a charged Brownian gas under the influence of external and non
uniform electric, magnetic and mechanical fields, immersed in a non uniform
bath temperature. With the collision time as an expansion parameter, we
study the solution to the associated Kramers equation, including a linear
reactive term. To first order we obtain the asymptotic (overdamped) regime,
governed by transport equations, namely: for the particle's density, a
Smoluchowski-reactive like equation; for the particle's momentum density, a
generalized Ohm's like equation; and for the particle\'s energy density, a
Maxwell-Cattaneo like equation. Defining a nonequilibrium temperature as the
mean kinetic energy density, and introducing Boltzmann's entropy density via
the one particle distribution function, we present a complete
thermohydrodynamical picture for a charged Brownian gas. We probe the
validity of the local equilibrium approximation, Onsager relations,
variational principles associated to the entropy production, and apply our
results to: carrier transport in semiconductors, hot carriers and Brownian
motors. Finally, we outline a method to incorporate non linear reactive
kinetics and a mean field approach to interacting Brownian particles.

\bigskip

\noindent Pacs {05.20.Dd, 05.40.-a, 05.40.Jc, 05.70.Ln, 51.10+y, 66.10.Cd,
82.40.-g}


\noindent Corresponding author: R. E. Lagos, monaco@rc.unesp.br

\vspace{2.0cm}

\begin{center}
{\large 1: Introduction }

\bigskip
\end{center}

The ubiquitous Brownian motion \cite{origin1}-\cite{origin8} remains an
outstanding paradigm in modern physics \cite{origin9}. The theoretical
framework \cite{gen01, gen02} consisting of Kramers equation (a
Fokker-Planck equation in phase space), Smoluchowski equation (asymptotic or
overdamped contraction of the latter) and the associated stochastic Langevin
equation, have been widely applied to diverse problems, such \ as: Brownian
motion in potential wells, chemical reactions rate theory, nuclear dynamics,
stochastic resonance, surface diffusion, general stochastic processes and
evolution of nonequilibrium systems, in both classical and quantum contexts.
More recent applications include thermodynamics of small systems, \
molecular motors, chemical and biological nanostructures, mesoscopic motors
power output and efficiency, and the fractional Kramers equation applied to
anomalous diffusion. Concerning reviews and applications, we mention some
representative but by no means an exhaustive list of references \cite{gen01}-%
\cite{gen49}, some monographs \cite{mon01}-\cite{mon10} in addition to the
"founding papers" \cite{origin3}-\cite{origin8}. In his celebrated 1943
Brownian motion paper \cite{gen02}, Chandrasekhar outlined the method for
solving a Brownian particle in a general field of force. It took
approximately sixty years to report exact solutions for the Brownian motion
of a charged particle in uniform and static electric and/or magnetic fields
\ \cite{mag01}-\cite{mag15}\ \ (see also some previous related works \cite%
{mag16,mag17}).

In Section 2 we generalize and extend previous work \cite{mag05, mag07} and
consider a charged Brownian particle in general field of force (including
magnetic fields), in an inhomogeneous medium (a nonuniform/nonisothermal
bath temperature profile \ \cite{temp00}-\cite{temp25}). Furthermore we
include a reactive linear term (BGK\ like \ \cite{bgk01}-\cite{bgk04}) and
obtain a general associated Kramers equation. In Section 3 we expand the
solution generalizing a previous result \cite{mag05} \ yielding recursive
relations (for other expansions in the literature see for example \cite%
{mon04,mag03} and \cite{exp1}-\cite{exp3}. In Section 4 we use these
recursive relations to obtain the Hydrothermodynamical picture of Brownian
motion. In Section 5 we present some straight forward applications namely,
the linear Onsager-like expansions for the entropy production for Brownian
motion, generalized Shockley carrier transport equations, hot carrier
transport and chemically reacting Brownian gases. Finally in Section \ 6 our
concluding remarks include further applications of our results.

\vspace{0.5cm}

\begin{center}
{\large 2: Generalized Kramers equation for a Brownian charged particle }

\vspace{0.5cm}
\end{center}

Consider a charged Brownian gas composed of $N$ particles (mass $m$, charge $%
q$) in an inhomogeneous bath temperature profile, under the influence of
external fields\ (not necessarily uniform in space, nor time independent).
Our starting point is \textbf{Kramers} equation for the one particle
distribution function $P(\mathbf{x},\mathbf{v,}t)$, at position $\mathbf{x}$%
, with $\mathbf{v}$\textbf{\ }and at time $t$, being in contact with an
inhomogeneous bath temperature $T(\mathbf{x)}$ (natural units $k_{B}=1$) and
under a general field of force: a mechanical contribution $\ \ \mathbf{F}_{%
\text{mec }}=-\mathbf{\nabla }W$, with $W=$ $-\mathbf{xF}_{\text{ext}}(t)+U_{%
\text{mec}}$ (allowing for potential derived forces and non potential
homogeneous external forces); an electric contribution $\mathbf{F}_{\text{%
elec}}=q\mathbf{E}=-q\mathbf{\nabla }\phi $ and a magnetic contribution
(Lorentz%
\'{}%
s velocity dependent force) $\mathbf{F}_{\text{mag}}\mathbf{=}\frac{1}{c}q%
\mathbf{v}\times \mathbf{B}$. The total external force can be casted as a
potential derived force plus a velocity contribution

\begin{equation}
\mathbf{F}(\mathbf{x,v},t)\mathbf{=\mathbf{\mathbf{F}}_{\text{mec}}\mathbf{(%
\mathbf{x},}}t\mathbf{\mathbf{)+\mathbf{F}}_{\text{elec}}\mathbf{(\mathbf{x},%
}}t\mathbf{\mathbf{)+\mathbf{F}}_{\text{mag}}\mathbf{(\mathbf{x},}}t\mathbf{%
\mathbf{)}=\mathbf{F}_{\text{pot}}-}m\mathbf{\omega \times v}
\end{equation}

where

\begin{eqnarray}
\mathbf{F}_{\text{pot}} &=&-\mathbf{\nabla }U  \notag \\
U &=&-\mathbf{xF}_{\text{ext}}(t)+U_{\text{mec}}+q\phi  \label{force} \\
\mathbf{\omega } &=&\frac{q}{mc}\mathbf{B}  \notag
\end{eqnarray}

\noindent\ \ \ For a charged Brownian particle, Kramers equation for the
distribution $P(\mathbf{x},\mathbf{v,}t)$ reads (see for example \cite%
{mag03,mag05, mag07})

\begin{equation}
\frac{\partial P}{\partial t}+\mathbf{v}\frac{\partial P}{\partial \mathbf{x}%
}+\frac{1}{m}\mathbf{F}(\mathbf{x},\mathbf{v,}t)\frac{\partial P}{\partial 
\mathbf{v}}=\frac{1}{\tau }\frac{\partial }{\partial \mathbf{v}}\left( 
\mathbf{v}P+\frac{T(\mathbf{x})}{m^{{}}}\frac{\partial P}{\partial \mathbf{v}%
}\right) =\frac{\partial }{\partial \mathbf{v}}\left( \tau ^{-1}\mathbf{v}%
P+\Gamma \frac{\partial P}{\partial \mathbf{v}}\right)  \label{kramers}
\end{equation}%
where the Brownian collision time is denoted by $\tau $, the mobility by $%
\lambda $, \ defined by the relation $m\lambda =$ $\tau $ ($\gamma =\lambda
^{-1}$ is Stokes friction coefficient). The mobility and the diffusion
coefficient $D$, are both related to the bath temperature via the celebrated
fluctuation dissipation theorem $\ D=\lambda k_{B}T$ (Sutherland-Einstein
relation \cite{origin4, origin5}). Notice that in our scheme the mobility
(or $\tau $) may also have a position dependent profile; and $\Gamma $ is
related to fluctuations, as given by $D=\tau ^{2}\Gamma .$ The left hand
side of the previous equation may be denoted as the streaming operator $%
\mathbb{L}$, operating on $P$ along the (Newtonian) \ trajectory $\mathbf{v=}%
\overset{\cdot }{\mathbf{x}}$, $\ m\overset{\cdot }{\mathbf{v}}=\mathbf{F}$.
The right hand side is denoted as the Fokker Planck collision kernel
operator $\mathbb{K}_{FP}$ where the first term $\ $is the contribution of
the (dissipative) Stokes force $\mathbf{F}_{\text{S}}=-\gamma \mathbf{v}$
and the second term the fluctuating contribution, so \ in explicit form we
have

\bigskip

\begin{equation}
\mathbb{L}(\mathbf{F)=}\frac{\partial }{\partial t}+\mathbf{v}\frac{\partial 
}{\partial \mathbf{x}}+\frac{1}{m}\mathbf{F}(\mathbf{x},\mathbf{v,}t)\frac{%
\partial }{\partial \mathbf{v}}
\end{equation}

\begin{equation}
\mathbb{K}_{FP}(\lambda ,\Gamma )=\frac{\partial }{\partial \mathbf{v}}%
\left( \tau ^{-1}\mathbf{v}+\Gamma \frac{\partial }{\partial \mathbf{v}}%
\right)
\end{equation}

and we may cast equation (\ref{kramers}) in the compact form

\bigskip

\begin{equation}
\mathbb{L}(\mathbf{F)}P(\mathbf{x},\mathbf{v,}t)=\mathbb{K}_{FP}(\tau
,\Gamma )P(\mathbf{x},\mathbf{v,}t)  \label{kramerc}
\end{equation}

Let us define a tensorial Stokes force \cite{mag05,mag07}\ as

\begin{equation}
\mathbf{F}_{TS}\mathbf{=-}\lambda ^{-1}\mathbf{v-}m\mathbf{\omega \times v=-M%
}^{-1}\mathbf{v}
\end{equation}

where the magneto mobility tensor $\mathbf{M}$\ is defined (when operating
over an arbitrary vector $\mathbf{V}$) as

\bigskip

\begin{equation}
\mathbf{M(}\tau \mathbf{,\omega )V}=\lambda \frac{\mathbf{V+}\tau \mathbf{V}%
\times \mathbf{\omega +}\tau ^{2}\mathbf{\omega }\left( \mathbf{\omega \cdot
V}\right) }{1+\tau ^{2}\mathbf{\omega }^{2}}  \label{tensor}
\end{equation}

\bigskip

In particular we notice the familiar form \cite{mag07} for the case $\mathbf{%
B=}B\widehat{z}$

\begin{equation}
\mathbf{M(}\tau \mathbf{,\omega )=}\frac{\lambda }{1+\tau ^{2}\omega ^{2}}%
\left( 
\begin{array}{ccc}
1 & \tau \omega & 0 \\ 
-\tau \omega & 1 & 0 \\ 
0 & 0 & 1+\tau ^{2}\omega ^{2}%
\end{array}%
\right)
\end{equation}

\bigskip It was shown \cite{mag04,mag07} \ that by defining such a tensorial
Stokes force equation (\ref{kramerc}) \ is equivalent to

\bigskip 
\begin{equation}
\mathbb{L}(U\mathbf{)}P(\mathbf{x},\mathbf{v,}t)=\mathbb{K}_{FP}(\mathbf{M,}%
\Gamma )P(\mathbf{x},\mathbf{v,}t)  \label{kramerf}
\end{equation}%
thus, the streaming operator includes only de (scalar) potential derived
force and the magnetic contribution enters as a dissipative non diagonal
contribution in the collision kernel (for convenience we have denoted $%
\mathbb{L}(U\mathbf{)=}\mathbb{L}(\mathbf{F}_{\text{pot}}=-\mathbf{\nabla }U%
\mathbf{)}$). Now as in \cite{mag05,bgk02} we add a BGK contribution to the
collision kernel denoted by $\mathbb{K}_{BGK}.$ This collision operator is
single relaxation time approximation \cite{bgk01}-\cite{bgk04} where the
distribution $P$ relaxes (collision time $\tau _{0}$) to a prescribed
distribution $P_{0}$, the latter instantly equilibrated in the velocity
coordinates, to the bath temperature, thus

\begin{equation}
P_{0}(\mathbf{x,v},t)=n_{0}(\mathbf{x},t)f_{0}(\mathbf{v},T(\mathbf{x}))%
\hspace{1cm}f_{0}(\mathbf{v,}m\mathbf{,}T(\mathbf{x}))=\left( \frac{m}{2\pi
T(\mathbf{x})}\right) ^{\frac{3}{2}}\exp \left( -\frac{m\mathbf{v}^{2}}{2T(%
\mathbf{x})}\right)
\end{equation}%
where $f_{0}$ is a (normalized) Maxwellian distribution, $n_{0}$ is a
prescribed density profile. For the Brownian gas we define particle density
as $\ n(\mathbf{x},t)=\int d\mathbf{v}P(\mathbf{x},\mathbf{v,}t)$\ \ where $%
N=\int d\mathbf{x}n(\mathbf{x,}t)$. In general \ \ $n_{0}(\mathbf{x},t)$
needs not to satisfy $N=\int d\mathbf{x}n_{0}(\mathbf{x,}t)$ \ (BGK is not a
particle conserving approximation). The BGK kernel is given by

\bigskip 
\begin{equation}
\mathbb{K}_{BGK}(n_{0}(\mathbf{x},t),\tau _{0},k)P(\mathbf{x},\mathbf{v,}t)=-%
\frac{1}{\tau _{0}}\left( P(\mathbf{x},\mathbf{v,}t)-P_{0}(\mathbf{x},%
\mathbf{v,}t)\right)
\end{equation}%
and represents the simple chemical reaction $A\longleftrightarrow A_{0}$,
both compounds with particle density $n(\mathbf{x},t)$ and $n_{0}(\mathbf{x}%
,t)$, respectively, and with forward (backward) rate $k_{+}(k_{-}),$ we have

\bigskip

\begin{equation}
\left( \frac{\partial n(\mathbf{x},t)}{\partial t}\right) _{\text{r}}=\int d%
\mathbf{v}\mathbb{K}_{BGK}P(\mathbf{x},\mathbf{v,}t)=-k_{+}n(\mathbf{x}%
,t)+k\_n_{0}(\mathbf{x},t)=-\left( \frac{\partial n_{0}(\mathbf{x},t)}{%
\partial t}\right) _{\text{r}}
\end{equation}%
where we have defined $\ $\ the rates as $k_{+}=$ $k_{-}=\tau _{0}^{-1}$.
Then, Kramers equation for a charged Brownian particles under a general
field of force in an inhomogeneous medium, including a linear reactive term,
is compacted as

\begin{equation}
\mathbb{L}(U\mathbf{)}P(\mathbf{x},\mathbf{v,}t)=\mathbb{K}_{FP}(\mathbf{M}%
,\Gamma )P(\mathbf{x},\mathbf{v,}t)+\mathbb{K}_{BGK}(n_{0}(\mathbf{x}%
,t),\tau _{0})P(\mathbf{x},\mathbf{v,}t)=\mathbb{K}P(\mathbf{x},\mathbf{v,}t)
\label{kram}
\end{equation}

This is a generalized kinetic equation in the Fokker-Planck (FP) context for
a charged Brownian particle. The\textbf{\ novel aspect in our approach} is
to include \textbf{on an equal footing} the following aspects: An external
magnetic field (a velocity dependent force ) rarely considered in the
Brownian context; an inhomogeneous bath temperature profile $T(\mathbf{x})$
and, in addition to the particle conserving FP collision kernel, we add a
non conserving particle collision contribution (a generalized BGK
mechanism), enabling our approach to incorporate chemical reactions and/or
generation-recombination processes (for the latter we may consider an
inhomogeneous collision time profile $\tau _{0}(\mathbf{x})$).

\vspace{0.5cm}

\newpage

\begin{center}
{\large 3: Recursive expansion of Kramers equation }

\bigskip
\end{center}

\noindent\ We expand Kramers equation (\ref{kram} ) in order to obtain a
recursive solution. Some expansion methods include \cite{exp1}-\cite{exp3},
and we review a method outlined in ref. \cite{mag05}. The expansion is
casted as

\begin{equation}
P(\mathbf{x},\mathbf{v,}t)=\sum_{n,m,l=0}^{\infty }\Psi _{nml}(\mathbf{x,}%
t)\Phi _{nml}(\mathbf{w})\hspace{1cm}\hspace{1cm}\mathbf{w=}\sqrt{\frac{1}{2}%
}\frac{\mathbf{v}}{v_{T}},\hspace{1cm}v_{T}^{2}=\frac{T(\mathbf{x})}{m}
\end{equation}

\noindent where $\Phi _{nml}(\mathbf{w)=}\phi _{n}(w_{x})\phi
_{m}(w_{y})\phi _{l}(w_{z})$ is the product of (normalized) Hermite
functions \cite{sneddon}. Following \cite{mag05} and using appropriate
recursion relations \cite{sneddon}, equation (\ref{kram}) reduces to a
difference-differential recursive system of equations for the $\Psi ^{\prime
}s$ \ ($\mathbf{v}$ has been integrated out). These recursion relations are
better displayed in the compact form, and as before \cite{mag05} we allow
for a slowly time varying (compared with $\tau $) external temperature field 
$T(\mathbf{x},t).$

\begin{equation}
k\frac{\tau }{\tau _{0}}n_{0}\delta _{\mathbf{n,0}}=\tau \mathbf{\omega
\cdot A}^{\ast }\mathbf{\times A}Z_{\mathbf{n}}+(\mathbf{A}^{\ast }\mathbf{A+%
}\tau R)Z_{\mathbf{n}}\hspace{1cm}\mathbf{n}=(n_{1},n_{2},n_{3})
\label{recu}
\end{equation}

\noindent where

\begin{equation}
R=\frac{\partial }{\partial t}+\frac{1}{\tau _{0}}+\mathbf{\nabla A^{\ast }+}%
v_{T}^{2}\left( \mathbf{\nabla A+A}\frac{\mathbf{\nabla }U}{T}\mathbf{+A}%
^{2}\left\{ (v_{T}^{2}\mathbf{A+A}^{\ast })\frac{\mathbf{\nabla }T}{2T}+%
\frac{1}{2T}\frac{\partial T}{\partial t}\right\} \right)
\end{equation}%
with $\mathbf{\nabla =}\frac{\partial }{\partial \mathbf{x}}$ and

\noindent

\begin{equation}
Z_{\mathbf{n}}=\frac{v_{T}^{n_{1}+n_{2}+n_{3}}\Psi _{n_{1}n_{2}n_{3}}}{\sqrt{%
n_{1}!n_{2}!n_{3}!}}  \label{coef}
\end{equation}

\noindent The asymmetric lowering and raising operators $\mathbf{A}$, $%
\mathbf{A}^{\ast }$ are defined respectively by

\begin{eqnarray}
A_{1}Z_{\mathbf{n}} &=&Z_{n_{1}-1,n_{2},n_{3}}\hspace{1cm}A_{1}^{\ast }Z_{%
\mathbf{n}}=(1+n_{1})Z_{n_{1}+1,n_{2},n_{3}}  \notag \\
A_{2}Z_{\mathbf{n}} &=&Z_{n_{1},n_{2}-1,n_{3}}\hspace{1cm}A_{2}^{\ast }Z_{%
\mathbf{n}}=(1+n_{2})Z_{n_{1},n_{2}+1,n_{3}} \\
A_{3}Z_{\mathbf{n}} &=&Z_{n_{1},n_{2},n_{3}-1}\hspace{1cm}A_{3}^{\ast }Z_{%
\mathbf{n}}=(1+n_{3})Z_{n_{1},n_{2},n_{3}+1}  \notag
\end{eqnarray}

\vspace{0.5cm}

\begin{center}
{\large 4: Hydrothermodynamics of Brownian Motion }

\vspace{0.5cm}
\end{center}

We now proceed in a standard fashion \cite{liboff}-\cite{kreuzer} and define
some relevant moments of the distribution $P(\mathbf{x},\mathbf{v,}t)$. The
usual (lower) moments are: particle density $n(\mathbf{x},t),$ particle flux
vector density $\mathbf{J}_{M}(\mathbf{x},t)$, pressure tensor density $%
\mathbf{\Pi }(\mathbf{x},t)$, kinetic energy density $E(\mathbf{x},t)$,
energy flux vector density $\mathbf{J}_{E}(\mathbf{x},t)$ and the total
energy flux vector density $\mathbf{J}_{Q}(\mathbf{x},t).$ Some useful
auxiliary quantities are also defined, namely: mass density $\rho (\mathbf{x}%
,t)$, mass flux density $\mathbf{J}_{\rho }(\mathbf{x},t)$, charge flux
density $\mathbf{J}_{q}(\mathbf{x},t)$, scalar pressure $p(\mathbf{x},t)$
and the stream velocity $\mathbf{u}(\mathbf{x},t)$. All of the above are
defined by:

\begin{equation}
n(\mathbf{x},t)=\int d\mathbf{v}P(\mathbf{x},\mathbf{v,}t),\hspace{1cm}\rho (%
\mathbf{x},t)=mn(\mathbf{x},t)
\end{equation}

\begin{equation}
\mathbf{J}_{M}(\mathbf{x,}t)=\int d\mathbf{vv}P(\mathbf{x},\mathbf{v,}t)=n(%
\mathbf{x},t)\mathbf{u}(\mathbf{x},t)
\end{equation}

\begin{equation}
\mathbf{J}_{\rho }(\mathbf{x,}t)=m\mathbf{J}_{M}(\mathbf{x,}t),\hspace{0.3cm}%
\mathbf{J}_{q}(\mathbf{x,}t)=q\mathbf{J}_{M}(\mathbf{x,}t)
\end{equation}

\begin{eqnarray}
\mathbf{\Pi }(\mathbf{x,}t) &=&m\int d\mathbf{vvv}P(\mathbf{x},\mathbf{v,}t),%
\hspace{1cm}p(\mathbf{x,}t)=\frac{1}{3}\sum_{i=1}^{3}\mathbf{\Pi }_{ii}(%
\mathbf{x,}t)  \notag \\
&& \\
\mathbf{\Pi }_{u}(\mathbf{x,}t) &=&m\int d\mathbf{v(v-u)(v-u)}P(\mathbf{x},%
\mathbf{v,}t)=\mathbf{\Pi }(\mathbf{x,}t)-\rho \mathbf{u(\mathbf{x,}t)u}(%
\mathbf{x,}t)  \notag
\end{eqnarray}

\begin{equation}
E(\mathbf{x,}t)=\frac{1}{2}m\int d\mathbf{v}\left\vert \mathbf{v}\right\vert
^{2}P(\mathbf{x},\mathbf{v,}t)=\frac{3}{2}p(\mathbf{x,}t)
\end{equation}

\begin{equation}
\mathbf{J}_{E}(\mathbf{x,}t)=\frac{1}{2}m\int d\mathbf{vv}\left\vert \mathbf{%
v}\right\vert ^{2}P(\mathbf{x},\mathbf{v,}t)
\end{equation}

\begin{equation}
\mathbf{J}_{Q}(\mathbf{x,}t)=\int d\mathbf{vv}\left( \frac{1}{2}m\left\vert 
\mathbf{v}\right\vert ^{2}+U(\mathbf{x},t)\right) P(\mathbf{x},\mathbf{v,}t)=%
\mathbf{J}_{E}(\mathbf{x,}t)+U(\mathbf{x},t)\mathbf{J}_{M}(\mathbf{x,}t)
\label{flowenergy}
\end{equation}

\bigskip

\bigskip These relevant moments can be readily associated with the expansion
coefficients $Z_{\mathbf{n}}$, see equation (\ref{coef})

\begin{equation}
n(\mathbf{x},t)=Z_{000},\hspace{0.5cm}\Phi _{000}(\mathbf{w)}=f_{0}(\mathbf{%
v,}m,T(\mathbf{x)})
\end{equation}

\bigskip

\begin{equation}
\mathbf{J}_{M}(\mathbf{x},t)=\left( 
\begin{array}{c}
Z_{100} \\ 
Z_{010} \\ 
Z_{001}%
\end{array}%
\right)
\end{equation}

\bigskip 
\begin{equation}
\mathbf{\Pi (\mathbf{x},}t\mathbf{)=}n\mathbf{(\mathbf{x},}t\mathbf{)}T%
\mathbf{(\mathbf{x})}+m\left( 
\begin{array}{ccc}
2Z_{200} & Z_{110} & Z_{101} \\ 
Z_{110} & 2Z_{020} & Z_{011} \\ 
Z_{101} & Z_{011} & Z_{002}%
\end{array}%
\right)
\end{equation}

\bigskip

\bigskip 
\begin{equation}
\mathbf{J}_{E}\mathbf{(\mathbf{x},}t\mathbf{)}=\frac{5}{2}T\mathbf{(\mathbf{x%
}})\mathbf{J}_{M}\mathbf{(\mathbf{x},}t\mathbf{)}+m\left( 
\begin{array}{c}
3Z_{300}+Z_{120}+Z_{102} \\ 
Z_{210}+3Z_{030}+Z_{012} \\ 
Z_{201}+Z_{021}+3Z_{003}%
\end{array}%
\right)
\end{equation}

\bigskip

\bigskip

Then, the recursive relations, equation (\ref{recu}), as outlined in \cite%
{mag05} can de compacted into a set of balance equations (conservation of
mass, momentum and energy, respectively) yielding the hydrodynamical
picture. For the particle density the recursive relations yields

\noindent

\begin{equation}
\frac{\partial n}{\partial t}+\mathbf{\nabla J}_{M}=-k_{+}n+k_{-}n_{0}=%
\left( \frac{\partial n}{\partial t}\right) _{\text{reac}}  \label{smolu}
\end{equation}

\noindent a \ generalized Smoluchowski equation with a dissipative
(reactive) term; \ for the particle flux we obtain

\bigskip 
\begin{equation}
\frac{\partial \mathbf{J}_{M}}{\partial t}+\left( \frac{1}{\tau }+\frac{1}{%
\tau _{0}}\right) \mathbf{J}_{M}=\mathbf{J}_{M}\times \mathbf{\omega -}\frac{%
n}{m}\mathbf{\nabla }U\mathbf{-}\frac{1}{m}\mathbf{\nabla \Pi }  \label{flow}
\end{equation}%
if casted in terms of the stream velocity, and using equation (\ref{smolu})
\ we obtain the Navier-Stokes equation for Brownian flow

\begin{equation}
\frac{\partial \mathbf{u}}{\partial t}+\left( \mathbf{u\nabla }\right) 
\mathbf{u+}\frac{1}{\rho }\mathbf{\nabla \Pi }_{u}=\frac{1}{m}\left( -%
\mathbf{\nabla }U+\mathbf{F}_{\text{diss}}\right) ,\hspace{1cm}\mathbf{F}_{%
\text{diss}}=-\left( \mathbf{\Lambda }^{-1}+\frac{mn_{0}}{\tau _{0}n}\right) 
\mathbf{u}  \label{navier}
\end{equation}%
other equivalent versions of equation (\ref{flow}) in terms of the charge
flux density and with $\mathbf{F}_{\text{ext }}=0$ \ in equation (\ref{force}%
) \ are

\bigskip

\begin{equation}
\tau ^{\ast }\frac{\partial \mathbf{J}_{q}}{\partial t}+\mathbf{J}%
_{q}=\sigma ^{\ast }\left( \mathbf{E}+R_{H}\mathbf{J}_{q}\times \mathbf{B}%
\right) \mathbf{-}\frac{q\tau ^{\ast }n}{m}\mathbf{\nabla }U_{\text{mec}}%
\mathbf{-}\frac{q\tau ^{\ast }}{m}\mathbf{\nabla \Pi }
\end{equation}

\begin{equation}
\mathbf{J}_{q}=\sigma ^{\ast }\mathbf{\Lambda }^{\times }\mathbf{(}\tau
^{\ast },\mathbf{\omega )}\left( \mathbf{E-}\frac{1}{q}\mathbf{\nabla }U_{%
\text{mec}}\mathbf{-}\frac{1}{qn}\mathbf{\nabla }\Pi \right) -m\mathbf{%
\Lambda (}\tau ^{\ast },\mathbf{\omega )}\frac{\partial \mathbf{J}_{q}}{%
\partial t}
\end{equation}

\noindent a generalized Ohm%
\'{}%
s law \cite{nicholson} where the effective relaxation time $\tau ^{\ast }$,
the effective conductivity $\sigma ^{\ast }$ and Hall$^{\prime }$s
coefficient are respectively given by

\begin{equation}
\frac{1}{\tau ^{\ast }}=\frac{1}{\tau }+\frac{1}{\tau _{0}},\hspace{0.5cm}%
\sigma ^{\ast }(\mathbf{x,}t)=\frac{q^{2}\tau ^{\ast }n}{m},\hspace{0.5cm}%
R_{H}(\mathbf{x,}t)=\frac{1}{nqc}
\end{equation}

\noindent\ \ \ 

Finally we obtain the energy balance equation

\begin{equation}
\frac{\partial E}{\partial t}+\mathbf{\nabla J}_{E}=-\mathbf{J}_{M}\mathbf{%
\nabla }U-\frac{1}{\tau _{0}}(E-E_{0})\hspace{1cm}E_{0}=\frac{3}{2}n_{0}T
\label{energy}
\end{equation}

Thermodynamics is introduced by defining the kinetic temperature $\Theta $ \ 
\cite{mag05,mag07,jou} (a generalization of the equipartition theorem,
assuming the Brownian particle to have only translational degrees of freedom
) defined as

\begin{equation}
E(\mathbf{x,}t)\equiv \frac{3}{2}n(\mathbf{x,}t)\Theta (\mathbf{x,}t)
\end{equation}%
when substituted into (\ref{energy}) yields

\begin{equation}
\frac{3}{2}n\frac{\partial \Theta }{\partial t}+\mathbf{\nabla J}_{Q}=\left( 
\frac{3}{2}\Theta +U\right) \mathbf{\nabla J}_{M}-\frac{3}{2}\frac{1}{\tau
_{0}}n_{0}(\Theta -T)  \label{energx}
\end{equation}

\noindent

We define entropy density $S$ and the entropy flux density $\mathbf{J}_{S}$
respectively as \cite{mag03,mag05,mag07} 
\begin{eqnarray}
S(\mathbf{x,}t) &=&-\int d\mathbf{v}P(\mathbf{x,v},t)\ln \kappa P(\mathbf{x,v%
},t)\hspace{1cm}  \notag \\
&&  \label{entropy} \\
\mathbf{J}_{S}(\mathbf{x,}t) &=&-\int d\mathbf{vv}P(\mathbf{x,v},t)\ln
\kappa P(\mathbf{x,v},t)  \notag
\end{eqnarray}

\noindent where the constant $\kappa $ ($\ln \kappa =-1+3\ln \frac{h}{m})$
is chosen such that under equilibrium conditions $S\rightarrow $ $S_{\text{eq%
}}$ \cite{mag05,kittel} (for a simple ideal gas, where $h$ is Planck's
constant).

\bigskip 
\begin{equation}
S_{\text{eq}}(n_{\text{eq}},T_{\text{eq}})=n_{\text{eq}}\left( \frac{5}{2}%
+\ln \frac{n_{Q}(T_{\text{eq}})}{n_{\text{eq}}}\right) ,\hspace{0.5cm}%
n_{Q}(T_{\text{eq}})=\left( \frac{2\pi m}{h^{2}}T_{\text{eq}}\right) ^{\frac{%
3}{2}}  \label{equentrop}
\end{equation}

\bigskip

An entropy balance equation is readily obtained \cite{huang}, yielding an
expression for the entropy production $\Sigma _{S}$ solely in terms of the
particle distribution

\begin{equation}
\frac{\partial S(\mathbf{x,}t)}{\partial t}+\frac{\partial \mathbf{J}_{S}(%
\mathbf{x,}t)}{\partial \mathbf{x}}=\Sigma _{S}=-\int d\mathbf{v(}1\mathbf{+}%
\ln \kappa P(\mathbf{x,v},t))\mathbb{K}(P)
\end{equation}

Finally \cite{mag03,mag05} we define the various thermodynamical \textbf{%
potential densities}, namely: $F$ (Helmholtz), $G$ (Gibbs), $\mu ^{I}$
(intrinsic chemical potential), and $\mu $ (total chemical potential),
respectively given by \cite{kittel}

\begin{eqnarray}
F(\mathbf{x,}t) &=&E(\mathbf{x,}t)-\Theta (\mathbf{x,}t)S(\mathbf{x,}t) 
\notag \\
&&  \notag \\
G(\mathbf{x,}t) &=&F(\mathbf{x,}t)+p(\mathbf{x,}t)=n(\mathbf{x,}t)\mu ^{I}(%
\mathbf{x,}t) \\
&&  \notag \\
\hspace{0.5cm}\mu (\mathbf{x,}t) &=&\mu ^{I}(\mathbf{x,}t)+U(\mathbf{x,}t) 
\notag
\end{eqnarray}

the equilibrium expression for the chemical potential is \cite{kittel}

\bigskip

\begin{equation}
\mu _{\text{eq}}(n_{\text{eq}},T_{\text{eq}})=-T_{\text{eq}}\ln \frac{%
n_{Q}(T_{\text{eq}})}{n_{\text{eq}}}
\end{equation}

All moments, or equivalently, the expansion coefficients $Z_{\mathbf{n}},$ $%
\mathbf{n}\neq \mathbf{0}$, see equation (\ref{coef}), are expandable in
powers of $\tau $, when solving the recurrence relation (\ref{recu}), and
considering $\tau \omega $ as an independent parameter. Hereafter, when
referring to expansion order, we mean in powers of $\tau $, for example, the
magneto mobility tensor $\mathbf{M}$\ in equation (\ref{tensor}), is a first
order quantity. We further expand the expansion coefficients as 
\begin{equation}
Z_{\mathbf{n}}=\sum_{k=1}^{\infty }\tau ^{k}Z_{\mathbf{n}}^{(k)}
\end{equation}

For $\left\vert \mathbf{\nabla }T\right\vert \neq 0$ we have $Z_{\mathbf{n}%
}^{(k)}\equiv 0$ for $k < integer\left[ \frac{N+2}{3}\right] $. For the
isothermal case $\left\vert \mathbf{\nabla }T\right\vert =\frac{\partial T}{%
\partial t}\equiv 0$ we have $Z_{\mathbf{n}}^{(k)}\equiv 0$ for $k<N.$

Now, we discuss several relevant cases concerning the heat bath interaction
with the Brownian gas (a gas of real particles or some collective property
of a given macroscopic system). \ In order to simplify the discussion we
denote the Brownian gas as the solute and the heat bath as the solvent (the
fluid in which the Brownian particle is immersed). As the solute fluctuates
and dissipates in the solvent, towards an equilibrium (or non equilibrium
steady) state, the solute temperature $\Theta (\mathbf{x},t)$ evolves
according to the Brownian dynamics (given by equation \ref{kramers})
adjusting itself to the solvent temperature $T(\mathbf{x,}t)$ and the
external fields acting on the solute. On the other hand the solvent
temperature is perturbed by the solute temperature, a very weak interaction
indeed, since as a heat bath, $T(\mathbf{x},t)$ relaxes instantly (or very
fast compared with $\tau $) to a prescribed heat bath temperature \ profile $%
T_{\text{in}}(\mathbf{x})$, the latter a stationary function to be
considered as both an initial and a boundary condition. We realize that $%
\mathbf{\nabla }T$ \ is not controlled at all in the interior of the system
but only on its surfaces. In the bulk, the solute performs its Brownian
evolution, while the solvent temperature relaxes (very fast) to the
prescribed profile. This is a unsatisfactory state of affairs since in the
bulk, the control parameter \ $T_{\text{in}}(\mathbf{x})$ evolves as $T(%
\mathbf{x,}t)$ perturbed by the solute, at temperature $\Theta (\mathbf{x,}%
t) $, which in turn evolves following $T(\mathbf{x,}t),$ thus a clear cut
distinction \ between controlled external (thermal) fields and the resulting
(thermal) fluxes cease to exist \cite{kreuzer}.

We analyze three simple cases, with no attempt to model the heat bath at
this point (see for example\cite{mon05,bath}).

- Strictly Isothermal Bath, with $T(\mathbf{x,}t)=T_{\text{in}}(\mathbf{x}%
)=T_{R}$

- Strictly Stationary Inhomogeneous Bath with $T(\mathbf{x,}t)=T_{\text{in}}(%
\mathbf{x}),$ $\left\vert \mathbf{\nabla }T\right\vert \neq 0$, and

- Weakly Interacting Bath, first order in $\tau ,$ where $T(\mathbf{x,}t)=T_{%
\text{in}}(\mathbf{x})$ only at the boundaries.

From the recursion relations (\ref{recu}), for the strictly isothermal case,
the solute temperature \ has a second order correction (in $\tau $), already
obtained by us, as an asymptotic solution to a particular isothermal exact
solution \cite{mag07}. We present our result, as in \cite{mag07}, in terms
of the stream velocity $\mathbf{u}(\mathbf{x},t)$, in turn expressed as a
magnetocovariant derivative

\bigskip 
\begin{equation}
\Theta (\mathbf{x},t)=T_{R}+\frac{1}{3}m\mathbf{u}^{2}(\mathbf{x},t)
\label{hotemp}
\end{equation}

\bigskip where

\begin{equation}
\mathbf{u}(\mathbf{x},t)=-\frac{T_{R}}{n(\mathbf{x},t)}\mathbf{M}\exp \left(
-\frac{U(\mathbf{x},t)}{T_{R}}\right) \mathbf{\nabla }\left[ n(\mathbf{x}%
,t)\exp \left( \frac{U(\mathbf{x},t)}{T_{R}}\right) \right]  \label{hotempu}
\end{equation}

This is a generalization of Shockley's expression for hot carriers effective
temperature \cite{hot, omar}. Shockley's result coincides with our result at
zero magnetic field and homogeneous carrier concentration (provide we
identify the hot carrier with the solute, at temperature $\Theta $, and the
solvent with the lattice, at temperature $T_{R}$).

For the weakly interacting bath, to first order in $\tau $, we obtain

\bigskip

\begin{equation}
\Theta (\mathbf{x,}t)=T(\mathbf{x,}t)+\frac{\tau }{2}\frac{\partial T(%
\mathbf{x,}t)}{\partial t}=T(\mathbf{x,}t+\frac{\tau }{2})  \label{telegtemp}
\end{equation}%
identical to equation (7.2) \ obtained by \cite{jou}, provided we identify
the solute with electrons, at temperature $\Theta $ and the solvent with the
lattice, at temperature $T.$

\vspace{0.5cm}

\begin{center}
\newpage

\bigskip {\large 5: Applications }

\bigskip
\end{center}

A)- For the strictly isothermal bath case, we present the second order
results, from the recursion relations (\ref{recu}) and by expanding (\ref%
{entropy}). Our results are equivalent to the asymptotic regime for an exact
solution \cite{mag07}: the local equilibrium approximation \cite{groot}-\cite%
{nicolis} is not fulfilled \cite{rubi}; in particular the entropy density
and the chemical potential are given, respectively by the expressions

\begin{eqnarray}
S(n(\mathbf{x},t),\Theta (\mathbf{x},t)) &=&S_{\text{eq}}(n(\mathbf{x}%
,t),\Theta (\mathbf{x},t))-\frac{m}{2T_{R}}n(\mathbf{x},t)\mathbf{u}^{2}(%
\mathbf{x},t)  \notag \\
&&  \label{nonleq} \\
\mu (n(\mathbf{x},t),\Theta (\mathbf{x},t)) &=&\mu _{\text{eq}}(n(\mathbf{x}%
,t),\Theta (\mathbf{x},t))+U(\mathbf{x},t)+\frac{1}{2}m\mathbf{u}^{2}(%
\mathbf{x},t)  \notag
\end{eqnarray}

where the temperature $\Theta $ and the stream \ \ velocity $\mathbf{u}$ are
given respectively by equations (\ref{hotemp}) and (\ref{hotempu}).

\begin{center}
\bigskip
\end{center}

B)- For the strictly stationary inhomogeneous bath case, the local
equilibrium approximation holds, \ in particular the entropy density and the
chemical potential are given, respectively by the expressions

\begin{eqnarray}
S(\mathbf{x},t) &=&S_{\text{eq}}\left( n(\mathbf{x},t),T_{\text{in}}(\mathbf{%
x})\right)  \notag \\
&& \\
\mu (\mathbf{x},t) &=&\mu _{\text{eq}}\left( n(\mathbf{x},t),T_{\text{in}}(%
\mathbf{x})\right) +U(\mathbf{x},t)  \notag
\end{eqnarray}

\bigskip

C)- For the weakly interacting bath, again an as in case A) the local
equilibrium approximation does not hold, to first order (in $\tau $). The
particle flux density is given by the (equivalent) expressions

\begin{eqnarray}
\mathbf{J}_{M}(\mathbf{x},t) &=&\mathbf{-}n(\mathbf{x},t)\mathbf{M\nabla }U(%
\mathbf{x},t)-\mathbf{M\nabla }\left( \Theta (\mathbf{x},t)n(\mathbf{x}%
,t)\right)  \notag \\
&&  \label{massflow} \\
&=&\mathbf{-}n(\mathbf{x},t)\mathbf{M\nabla (}U(\mathbf{x},t)+\Theta (%
\mathbf{x},t))-\mathbf{D(\mathbf{x},}t\mathbf{)\nabla }n(\mathbf{x},t),%
\hspace{0.5cm}  \notag
\end{eqnarray}

where $\mathbf{D}=\Theta (\mathbf{x},t)\mathbf{M}$ is the magneto-diffusion
tensor. The total energy flux density is given by

\begin{equation}
\mathbf{J}_{Q}(\mathbf{x},t)\mathbf{=}\left( \frac{5}{2}\Theta (\mathbf{x}%
,t)+U(\mathbf{x},t)\right) \mathbf{J}_{M}(\mathbf{x},t)-\frac{5}{6}n(\mathbf{%
x},t)\Theta (\mathbf{x},t)\mathbf{M}_{3}\mathbf{\nabla \Theta }(\mathbf{x},t)
\label{heatflow}
\end{equation}

and where we have incorporated, at no cost, an inhomogeneous collision time
profile $\tau (\mathbf{x})$ as well as a space varying magnetic field, with
the compact definitions

\begin{eqnarray}
\mathbf{M}_{k}\mathbf{V} &=&\theta (\mathbf{x})\left( \mathbf{V+}\tau _{k}(%
\mathbf{x)V}\times \mathbf{\omega (\mathbf{x)}+}\tau _{k}^{2}(\mathbf{%
x)\omega }(\mathbf{x)}\left( \mathbf{\omega (\mathbf{x)}\cdot V}\right)
\right)  \notag \\
\hspace{1cm}\theta (\mathbf{x}) &=&\frac{\lambda (\mathbf{x)}}{1+\left( \tau
_{k}(\mathbf{x)\omega (x)}\right) ^{2}}\hspace{1cm}\tau _{k}(\mathbf{x)=}%
\frac{1}{k}\tau (\mathbf{x)} \\
\mathbf{M}_{k}\mathbf{(}\tau \mathbf{,\omega )} &\mathbf{=}&\mathbf{M(}\frac{%
\tau }{k}\mathbf{,\omega )\hspace{1cm}M}_{1}=\mathbf{M\hspace{1cm}}k=1,2,....
\notag
\end{eqnarray}

Notice that in equations (\ref{massflow}) and (\ref{heatflow}) \ \ according
to equation (\ref{telegtemp}) $\Theta (\mathbf{x},t)$ can be substituted by $%
T(\mathbf{x},t)$ since the magnetic mobility tensor $\mathbf{M}$, equation (%
\ref{tensor}), is already a first order quantity. With the last observation
in mind, we rewrite the \textquotedblleft fluxes\textquotedblright\ in terms
of the \textquotedblleft forces\textquotedblright , following \cite%
{kreuzer,desloge}. Thus, we define the matter and heat \textquotedblleft
forces\textquotedblright\ as

\begin{equation}
\mathbf{X}_{M}(\mathbf{x,}t)=-\nabla \left( \frac{\mu }{T}\right) \hspace{1cm%
}\mathbf{X}_{Q}(\mathbf{x,}t)=\nabla \left( \frac{1}{T}\right)
\end{equation}%
where

\begin{equation}
\mu (\mathbf{x},t)=\mu _{\text{eq}}\left( n(\mathbf{x},t),T(\mathbf{x}%
,t)\right) +U(\mathbf{x},t)
\end{equation}%
so to first order in $\tau $ we have the linear flux-force relations

\begin{eqnarray}
\mathbf{J}_{M}(\mathbf{x,}t) &=&\mathbf{L}_{MM}(\mathbf{x,}t)\mathbf{X}_{M}(%
\mathbf{x,}t)+\mathbf{L}_{MQ}(\mathbf{x,}t)\mathbf{X}_{Q}(\mathbf{x,}t) 
\notag \\
&& \\
\mathbf{J}_{Q}(\mathbf{x,}t) &=&\mathbf{L}_{QM}(\mathbf{x,}t)\mathbf{X}_{M}(%
\mathbf{x,}t)+\mathbf{L}_{QQ}(\mathbf{x,}t)\mathbf{X}_{Q}(\mathbf{x,}t) 
\notag
\end{eqnarray}%
where the coefficients are given by

\begin{eqnarray}
\mathbf{L}_{MM}(\mathbf{x,}t) &=&n(\mathbf{x,}t)T(\mathbf{x,}t)\mathbf{M} 
\notag \\
\mathbf{L}_{MQ}(\mathbf{x,}t) &=&\mathbf{L}_{QM}(\mathbf{x,}t)=n(\mathbf{x,}%
t)T(\mathbf{x,}t)\left( \frac{5}{2}T(\mathbf{x,}t)+U(\mathbf{x,}t)\right) 
\mathbf{M}  \label{onsager} \\
\mathbf{L}_{QQ}(\mathbf{x,}t) &=&n(\mathbf{x,}t)T(\mathbf{x,}t)\left( \frac{5%
}{2}T(\mathbf{x,}t)+U(\mathbf{x,}t)\right) ^{2}\mathbf{M+}\frac{5}{2}n(%
\mathbf{x,}t)T^{3}(\mathbf{x,}t)\mathbf{M}  \notag
\end{eqnarray}

\begin{equation}
\mathbf{L}_{QM}^{\dagger }(\mathbf{B})=\mathbf{L}_{MQ}^{{}}(-\mathbf{B})
\end{equation}%
the last equation indicating that Onsager relations are satisfied \cite%
{kreuzer}, \cite{groot}-\cite{nicolis}. Nevertheless, the Onsager like
coefficients do \ not depend solely on the equilibrium values of the state
variables (mass density, temperature). Instead they are functions of the
external potentials and nonequilibrium state variables, namely $n(\mathbf{x,}%
t)$ and $\Theta (\mathbf{x,}t)$, which in turn evolve according to equations
(\ref{smolu}) and (\ref{energx}) respectively (for simplicity of
presentation we omit the spatial and temporal dependence)

\begin{eqnarray}
\frac{\partial n}{\partial t} &=&-\mathbf{\nabla J}_{M}-k_{+}n+k_{-}n_{0} 
\notag \\
&& \\
\frac{3}{2}n\frac{\partial \Theta }{\partial t} &=&-\left( \frac{5}{2}%
\mathbf{\nabla }\Theta +\mathbf{\nabla }U\right) \mathbf{J}_{M}-\Theta 
\mathbf{\nabla J}_{M}+\frac{5}{6}\mathbf{\nabla }\left( n\Theta \mathbf{M}%
_{3}\mathbf{\nabla }T\right) -\frac{3}{2}\frac{1}{\tau _{0}}n_{0}(\Theta -T)
\notag
\end{eqnarray}%
and substituting from equation (\ref{telegtemp}), within the linear
approximation (first order in $\tau $) we obtain the coupled equations for
the evolution of the state variables $n(\mathbf{x,}t)$ and $T(\mathbf{x,}t)$
(or $\Theta $)

\bigskip

\begin{equation}
\frac{\partial n}{\partial t}+\frac{1}{\tau _{0}}(n-n_{0})=\mathbf{\nabla M}%
_{1}\mathbf{Y\hspace{1cm}Y=}n\mathbf{\nabla }U+n\mathbf{\nabla }T+T\mathbf{%
\nabla }n  \label{smolureact}
\end{equation}

\begin{equation}
\frac{3}{2}\left( n+\frac{\tau n_{0}}{2\tau _{0}}\right) \frac{\partial T}{%
\partial t}+\frac{3\tau n}{4}\frac{\partial ^{2}T}{\partial t^{2}}=\left( 
\frac{5}{2}\mathbf{\nabla }T+\mathbf{\nabla }U\right) \mathbf{M}_{1}\mathbf{Y%
}+T\mathbf{\nabla (\mathbf{M}}_{1}\mathbf{Y)+}\frac{5}{6}\mathbf{\nabla }%
\left( \mathbf{M}_{3}nT\mathbf{\nabla }T\right)  \label{cataheat}
\end{equation}

Equation (\ref{smolureact}) is a generalized Smoluchowski like advection
diffusion reactive equation, and equation (\ref{cataheat}) is a
Maxwell-Cattaneo like equation, generalizing Fourier's heat equation,
incorporates \ inertial effects \cite{mon10,jou},\cite{cata1}-\cite{cata3},
as noticed from the time delay character of equation (\ref{telegtemp}),
where the time needed for the acceleration of the heat flow is considered.

Thus, even though Onsager relations are satisfied, the coefficient's
dependence on nonequilibrium state variables and external fields, indicates
that any variational principle associated with the entropy production \ \ 
\cite{kondepudi, glansdorff,nicolis, luzzi1,luzzi2}, is not applicable to
the present case \cite{mon09,vlad}

\bigskip

In the absence of the BGK reactive term, null magnetic field, and for the
strictly stationary inhomogeneous bath case, we recover the inhomogeneous
media advection-diffusion equation, presented by \cite{temp08} \ and given by

\begin{equation}
\frac{\partial n(\mathbf{x,}t)}{\partial t}=\mathbf{\nabla }\left( \lambda (%
\mathbf{x)}\left[ n(\mathbf{x,}t)\mathbf{\nabla }U(\mathbf{x})+n(\mathbf{x,}%
t)\mathbf{\nabla }T(\mathbf{x})+T(\mathbf{x})\mathbf{\nabla }n(\mathbf{x,}t)%
\right] \right)
\end{equation}%
and for the rigid conductor (constant particle density $n$), no reactive
term and null external fields, we retrieve the inertial corrections to
Fourier's law, known as the Maxwell-Cattaneo equation \cite{cata1}-\cite%
{cata3} (of the hyperbolic type, known also as the telegrapher's equation)

\begin{equation}
\frac{\partial T(\mathbf{x,}t)}{\partial t}+\frac{\tau }{2}\frac{\partial
^{2}T(\mathbf{x,}t)}{\partial t^{2}}=K\left( T(\mathbf{x,}t)\mathbf{\nabla }%
^{2}T(\mathbf{x,}t)+\left( \mathbf{\nabla }T(\mathbf{x,}t)\right)
^{2}\right) =-\mathbf{\nabla J}_{F}(T(\mathbf{x,}t))
\end{equation}

\begin{equation}
\mathbf{J}_{F}(T)=KT^{3}(\mathbf{x,}t)\mathbf{\nabla }\left( \frac{1}{T(%
\mathbf{x,}t)}\right) =-KT(\mathbf{x,}t)\mathbf{\nabla }T(\mathbf{x,}%
t)=-\kappa (\mathbf{x,}t)\nabla T(\mathbf{x,}t),\hspace{0.5cm}
\end{equation}%
with $K=\frac{20}{9}\lambda .$The right hand side of the last equation is
known as the fundamental form of Fourier's law \cite{mon03}

\bigskip

D)-Carrier transport in semiconductors, in the Brownian scheme, for the
strictly stationary inhomogeneous bath case \cite{ashcroft,shockley}.
Consider two Brownian gases, say electrons (density $n_{\text{c}}$,
collision time $\tau _{\text{c}}$, mobility $\mathbf{M}_{\text{c}}=\mathbf{M}%
(\tau _{\text{c}}$,$\omega _{\text{c}})$, effective mass $m_{\text{c}}$ and
charge $-e$) and holes (density $p_{_{\text{v}}}$, collision time $\tau _{%
\text{v}}$, mobility $\mathbf{M}_{\text{v}}=\mathbf{M}(\tau _{\text{v}}$,$%
\omega _{\text{v}})$, effective mass $m_{\text{v}}$ and charge $e$, and with 
$m_{\text{c,v }}c=\omega _{\text{c,v}}|e\mathbf{B}|$).The BGK mechanism is
associated in this case to carrier generation-recombination with the
respective generation-recombination times $\tau _{c,v}^{0}$ and with the $%
n_{{}}^{0}$'s the respective equilibrium carrier concentrations. Then, by
applying equation (\ref{smolureact}) we obtain

\begin{eqnarray}
\frac{\partial n_{\text{c}}}{\partial t} &=&\mathbf{\nabla M}_{\text{c}%
}\left( -en_{\text{c}}\mathbf{E+}n_{\text{c}}\mathbf{\nabla }T+T\mathbf{%
\nabla }n_{\text{c}}\right) -\frac{1}{\tau _{\text{c}}^{0}}\left( n_{\text{c}%
}-n_{\text{c}}^{0}\right)  \notag \\
&& \\
\frac{\partial p_{\text{v}}}{\partial t} &=&\mathbf{\nabla M}_{\text{v}%
}\left( +ep_{\text{v}}\mathbf{E+}p_{\text{v}}\mathbf{\nabla }T+T\mathbf{%
\nabla }p_{\text{v}}\right) -\frac{1}{\tau _{\text{v}}^{0}}\left( p_{\text{v}%
}-p_{\text{v}}^{0}\right)  \notag
\end{eqnarray}

Shockley's equations (unidimensional, isothermal, null magnetic field case)
are readily recovered \cite{ashcroft,shockley}. Inhomogeneous
generalizations of Shockley's equations are of current interest \cite{semi1}-%
\cite{semi8}.

\bigskip

E)-Our scheme can readily be generalized for several Brownian gases ($\alpha
=1,...,l$) interacting via non reactive potential $U_{\alpha \beta }$ and
with the usual BGK reactive term associated to each gas. By applying a mean
field approximation scheme to a many component Brownian gas \cite{aguirre1,
aguirre2}, as worked out in several contexts in \cite{hydroint}-\cite%
{robcpa2}, Kramers system of equations can readily be generalized to

\begin{equation}
\mathbb{L}_{\alpha }(U_{\alpha }^{\text{eff}}\mathbf{)}P_{\alpha }(\mathbf{x}%
,\mathbf{v,}t)=\left( \mathbb{K}_{FP}(\mathbf{M}_{\alpha }\mathbf{,\Gamma }%
_{\alpha })+\mathbb{K}_{BGK}^{\alpha }\right) P_{\alpha }(\mathbf{x},\mathbf{%
v,}t)
\end{equation}%
where

\begin{equation}
U_{\alpha }^{\text{eff}}(\mathbf{x},t)=U_{\alpha }(\mathbf{x},t)+\sum_{\beta
}\int d\mathbf{y}U_{\alpha \beta }(\mathbf{x-y)}n_{\beta }(\mathbf{y},t)%
\hspace{1cm}n_{\beta }(\mathbf{y},t)=\int d\mathbf{v}P_{\beta }(\mathbf{y},%
\mathbf{v,}t)
\end{equation}

\bigskip

F)- We briefly sketch how to incorporate several chemical reactions schemes,
beyond the BGK scheme (deterministic, stochastic, photoreactions, etc., see
for example \cite{robco1}-\cite{robsolar}) in order to model molecular
motors. We present two chemical reactions that can be approximated at the
Kramers equation level, yielding the correct Smoluchowski- reactive equation.

For the nonlinear reaction \ \ 

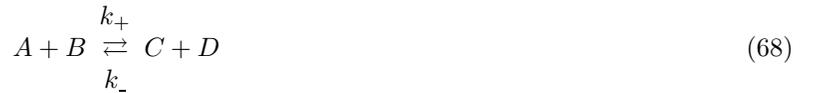
\begin{equation}
A+B%
\begin{array}{c}
k_{+} \\ 
\rightleftarrows \\ 
k_{\_}%
\end{array}%
C+D
\end{equation}

the kinetic equations are retrieved if we  consider the BGK like scheme

\begin{eqnarray}
\mathbb{K}_{BGK}^{A}P_{A}(\mathbf{x},\mathbf{v,}t)
&=&-k_{+}n_{B}P_{A}+k_{-}n_{C}n_{D}f_{0}(\mathbf{v,}m_{A},T)  \notag \\
&&  \notag \\
\mathbb{K}_{BGK}^{B}P_{B}(\mathbf{x},\mathbf{v,}t)
&=&-k_{+}n_{A}P_{B}+k_{-}n_{C}n_{D}f_{0}(\mathbf{v,}m_{B},T)  \notag \\
&& \\
\mathbb{K}_{BGK}^{C}P_{C}(\mathbf{x},\mathbf{v,}t)
&=&-k_{-}n_{D}P_{C}+k_{+}n_{A}n_{B}f_{0}(\mathbf{v,}m_{C},T)  \notag \\
&&  \notag \\
\mathbb{K}_{BGK}^{D}P_{D}(\mathbf{x},\mathbf{v,}t)
&=&-k_{-}n_{C}P_{D}+k_{+}n_{A}n_{B}f_{0}(\mathbf{v,}m_{D},T)  \notag
\end{eqnarray}

\bigskip

Finally, for the reaction cycle

\begin{equation}
A_{1}\overset{k_{1}}{\longrightarrow }A_{2}\overset{k_{2}}{\longrightarrow }%
A_{3}\overset{k_{3}}{\longrightarrow }A_{4}\cdot \cdot \cdot A_{l}\overset{%
k_{l}}{\longrightarrow }A_{1}
\end{equation}

we reproduce the correct kinetics with the approximation

\begin{equation}
\mathbb{K}_{BGK}^{\alpha }P_{\alpha }(\mathbf{x},\mathbf{v,}t)=-k_{\alpha
}P_{\alpha }+k_{\alpha -1}f_{0}^{(\alpha )}(\mathbf{v},m_{\alpha
},T)n_{\alpha -1}\hspace{1cm}n_{0}=n_{l}\hspace{1cm}\alpha =1,...l
\end{equation}

\bigskip

\begin{center}
{\large 6: Concluding remarks }

\bigskip\ \ 
\end{center}

We have extended our previous work on charged Brownian particles \cite%
{mag03,mag05,mag07}, in order to obtain a consistent expansion scheme in
powers of the collision time. We presented the complete hydrothermodynamical
picture for charged Brownian particles in the Kramers equation scheme,
considering the action of external magnetic, electric and mechanical fields,
chemical transformations within the BGK scheme, and furthermore space
dependent thermal fields (inhomogeneous media).

We developed a recursive method, in order to expand in powers of the
collision time $\tau $, the several moments of the solution of Kramers
equation, enabling us to compute the governing equations for mass (charge),
momentum (stream velocity) and energy (heat) \ flow. In the appropriate
limits we retrieve previous results, and present novel formulations and
results with immediate physical consequences.

Applications A, B and C (previous section) clearly indicates that the
Brownian scheme cannot be modelled, in general and in earnest, by the local
equilibrium approximation, no variational principle associated with entropy
production is applicable, and hyperbolic extensions of Fourier heat
conduction law must be considered even in lowest order. \ Therefore,
variational principles relevant to the Brownian scheme, must go beyond
entropy production \ considerations, and include among others, free
energies, efficiency factors and power output as well \cite{mon09,mon10} \
and connected to nonequilibrium fluctuation and work relations \ \cite%
{fluct1}-\cite{fluct10}.

In the following Applications items we briefly outline, from the formalism
developed in this work, some novel schemes: D) the incorporation of magnetic
field effects on the nonequilibrium state variables (including an
application for carrier transport in inhomogeneous semiconductors); E)
inclusion of interacting Brownian fluids via mean field approximation
schemes, and F) inclusion of chemical reactions to Kramer's approach to
Brownian motion.

Our future work concentrates in variational principles associated to the
Brownian motion, inclusion of chemical reactions and the magnetic field
effects in hot carrier transport in semiconductors \cite{gaucho}.

\begin{center}
\bigskip

\bigskip

\bigskip\ \ \ \ 
\end{center}

\noindent \emph{Acknowledgment}: This work was partially supported by CNPq
(Brasil).

\bigskip

\end{document}